\documentclass[showpacs,aps,prl,amsmath,amssymb,twocolumn]{revtex4-1}
\usepackage{graphicx}
\usepackage{dcolumn}
\usepackage{bm,times,color}
\usepackage{txfonts}
\usepackage{amsmath,epsfig}
\usepackage{epstopdf}
\newcommand{\ket}[1]{\left\vert#1\right\rangle}

\newcommand{\eq}{Eq.~}

\newcommand{\ug}{\!=\!}
\newcommand{\fig}{Fig.~}

\newcommand{\up} {\uparrow}
\newcommand{\down} {\downarrow}

\begin{document}

\author{F. Ciccarello$^1$, M. Paternostro$^2$, S. Bose$^3$, D. E. Browne$^3$, G. M. Palma$^4$, and M. Zarcone$^1$}
\affiliation{$^1$
CNISM and Dipartimento di Fisica e Tecnologie Relative, 
Universita' degli Studi di Palermo, Viale delle Scienze, Edificio 18, I-90128 Palermo, Italy\\
$^2$School of Mathematics and Physics, Queen's University, Belfast BT7 1NN, United Kingdom\\
$^3$Department of Physics and Astronomy, University College London, Gower Street, London WC1E 6BT, United Kingdom\\
$^4$ NANO - Istituto Nanoscienze - CNR and Dipartimento di Scienze Fisiche ed Astronomiche, 
Universita' degli Studi di Palermo, Via Archirafi 36, I-90123 Palermo, Italy}


\title{Physical model for the generation of ideal resources in multipartite quantum networking}

\date{\today}
\begin{abstract}
We propose a physical model for generating multipartite entangled states of spin-$s$ particles that have important applications in distributed quantum information processing. Our protocol is based on a process where mobile spins induce the interaction among remote scattering centers. As such, a major advantage lies on the management of stationary and well separated spins. Among the generable states, there is a class of $N$-qubit singlets allowing for optimal quantum telecloning in a scalable and controllable way. We also show how to prepare Aharonov, W and Greenberger-Horne-Zeilinger states.
\end{abstract}
\maketitle

\noindent
Multipartite entanglement is a near universal feature of coherently
interacting many-particle quantum systems. While, previously, multipartite
entanglement was at best a curiosity and, at worst, an unfortunate
impediment to the efficient classical simulation of the behavior of a system, the
development of quantum information science has made clear that certain
entangled states are valuable resources for information processing
and communication tasks. Although some recent studies have shown that not 
any entangled state is useful for the efficient processing of quantum information~\cite{bipartite,multipartite}, multipartite entanglement clearly plays a prominent role in a distributed quantum information processing
scenario \cite{GHZ,Browne}.

Here we propose a strategy for the generation of a variety of multipartite entangled states that have indeed relevant applications in quantum communication. Our scheme is based on  {spin symmetries} and exploits a scattering-based mechanism, involving well separated standing spins, that holds clear promises for  implementation in spintronics and quantum optics setups~\cite{prl,mappaNP}. It delivers a variety of multi-spin states that are ideal resources for protocols such as the Byzantine agreement~\cite{gisin}, secret sharing~\cite{hillery} and  telecloning~\cite{murao}. In particular, we set a physical model for the systematic and scalable preparation of classes of states that, despite playing a key role in the implementation of a few quantum protocols, have so far been realized only via clever experimental craftsmanship~\cite{qss,munich,MauroDicke}. Our scheme, which is very flexible and able to generate a variety of multipartite states, would pave the way to the actual implementation of multi-party quantum protocols for communication (a major step forward in its own right) and the study of the physical properties of interesting classes of multipartite entangled states.

\begin{figure}[b]
\center{\psfig{figure=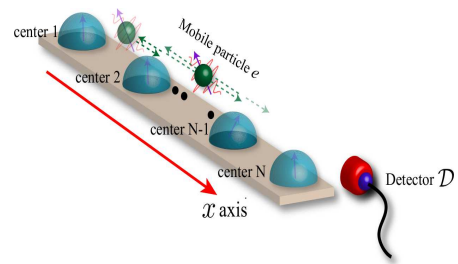,width=5.85cm,height=3.5cm}}
\caption{(Color online) Sketch of the physical proposal. Unpolarized mobile particles $e$'s undergo scattering from a linear array of $N$ static spins. When these are spaced in a way so as to satisfy the resonance conditions of the problem and initially prepared in a suitable product state, successive records of the mobile particles in the transmission channel extract an $N$-partite singlet state of the scattering centers.}
\label{Super}
\end{figure} 
The set-up we consider is sketched in Fig.~\ref{Super}. It consists of a mobile spin-1/2 particle $e$ propagating across a one-dimensional wire (aligned along the $x$-axis of a reference frame). At positions $x{=}x_i$ ($i{=}1,2,..,N$), the spin degree of freedom of $e$ interacts with that of a spin-$s$ scattering center $i$ according to the Hamiltonian  $\hat{H}{=}\hat{H}_{kin}{+}\hat{V}$ with (throughout this work we set $\hbar{=}1$)
\begin{equation} 
\label{Vel}
\hat{V}{=} J\sum^N_{i=1}\hat{\mbox{\boldmath$\sigma$}}\cdot \hat{\mathbf{S}}_i\,\delta(x-x_i).
\end{equation}
Here, $\hat{\mbox{\boldmath$\sigma$}}$ and $\hat{\mathbf{S}}_i$ are the (pseudo) spin operators of $e$ and the $i^\text{th}$ center respectively, $\hat{H}_{kin}$ is the kinetic energy of $e$ and $J$ is the {contact spin-spin} interaction strength~\cite{notaXY}. 
We {assume that the wavevector $k$ of a mobile particle fulfills the {\it resonance conditions} (RCs) $k (x_{j+1}{-}x_j){=}q_j \pi$ ($q_j\in\mathbb{Z}$, $j{=}1,..,N\!-\!1$) and we do not require any knowledge on the state of $e$, which could be even prepared in a maximally-mixed spin state. In Refs.~\cite{NJP,mappaNP} it was shown that, under such working conditions, the dynamics takes place as if the static particles all occupy the same site,
{\it i.e.} $\delta_{RC}(x-x_j){=}\delta_{RC}(x-x_1)~\forall{j}{=}1,..,N$ (where the subscript RC reminds that resonance conditions are set)~\cite{nota1}. This leaves us with the effective representation of the scattering potential~\cite{mappaNP}
\begin{equation} 
\label{Veff-el-N}
\hat{V}_{RC}{=} 
J\,\hat{\mbox{\boldmath$\sigma$}}\cdot \hat{\mathbf{S}}\,\delta_{RC}(x)\,,
\end{equation}
where we have set $x_1{=}0$ and $\mathbf{\hat{S}}{=} \sum_{i{=}1}^{N}\hat{\mathbf{S}}_i$ is the total spin of the scattering centers. \eq(\ref{Veff-el-N}) shows that under RCs, due to $[\mathbf{\hat{S}}^2, \hat{V}_{RC}]{=}0$,  $\mathbf{\hat{S}}^2$ is a constant of motion. Here, we indicate with $|\Psi^-_{s,N}\rangle$ the singlet state of $N$ scattering centers of spin quantum number $s$. In the case of $N{=}2$ and for an unpolarized spin state of $e$, it is easily checked \cite{mappaNP, jpa} that the two-center singlet state $|\Psi^-_{s,2}\rangle_{12}$ is the only one such that the effective representation of $\hat{V}$ in Eq.~(\ref{Veff-el-N}) vanishes. We recall that, for $s{=}1/2$, $|\Psi^-_{1/2,2}\rangle{=}(|{\up\down}\rangle{-}|{\down\up}\rangle)/\sqrt{2}$ with $\{|{\up}\rangle,|{\down}\rangle\}$ indicating single-spin states. Conversely, when $\hat{V}_{RC}{=}0$, particles $1$ and $2$ are necessarily in the singlet state. This implies that under RCs and when the centers are in $|\Psi^-_{1/2,2}\rangle$, an injected particle $e$ will tunnel across the interaction region with unit probability. Hence, in an experiment based on the mechanism above, if a large enough number of mobile particles are transmitted across the interaction region, we can conclude that an effective projection over $|\Psi^-_{s,2}\rangle$ has been performed \cite{mappaNP, yuasa} (see Fig.~\ref{Super}). 

For initially uncorrelated scattering centers with $s{=}1/2$, this occurs with a maximum success probability equal to $1/2$ when particles $1$ and $2$ are prepared in $|{\uparrow\,\! \downarrow}\rangle$ or $|{\downarrow\,\! \uparrow}\rangle$, which are states having maximum overlap with $|\Psi^-_{1/2,2}\rangle$. Similar conclusions can be drawn for higher-$s$ particles \cite{prl, mappaNP}. The scheme enjoys many interesting features, such as robustness against static disorder, imperfect arrangements of RCs and phase noise~\cite{robust, mappaNP}. Also, it does not require fine control of the interaction times/coupling strengths nor the preparation/post-selection of  the $e$'s spin~\cite{mappaNP, yuasa} .  

\noindent
{\it Occurrence and extraction of multipartite singlet states.--}  
{In what follows, we prove that the {singlet-state extraction} protocol is effective also for an \textit{arbitrary number} $N$ of scattering centers}. In general, we can state that \emph{if} an $N$-party singlet state exists then under RCs such state ensures ${\hat{V}_{RC}{=}}0$, in a way {analogous} to the two-center case. However, the availability and number of singlet states for a given configuration crucially depends on the number of scattering centers $N$ and their spin quantum number $s$. Indeed, by using standard addition of angular momenta one can see that no singlet state occurs when $s$ is half-integer and $N$ is odd, while {\emph{many} singlet states are in order for even $N$ and given values of $s$}. 

Before proceeding further we remind that when more than two angular momenta are added together, several nonequivalent coupling schemes can be used to build up the coupled basis~\cite{aam}, {\it i.e.} the common eigenstates of $\hat{\mathbf{S}}^2$ and $\hat{S}_z$ whose quantum numbers are denoted as $S$ and $M$, respectively. We label $s_i$ ($i{=}1,..,N$) the quantum number associated with $\hat{\mathbf{S}}_i^2$, which is the squared total spin of the $i^\text{th}$ center. For $K\!< \!N$ centers $\{i_1$, $i_2$,..,$i_K\}$, we call $s_{i_{1}i_{2}\cdot\cdot\cdot i_{K}}$ the quantum number associated with $\sum^K_{j=1}\hat{\mathbf{S}}^2_{i_j}$ (which becomes the squared total spin $\mathbf{S}^2$ when $K{=}{N}$). To illustrate the role played by the symmetries of the setting in \fig1, we {first} discuss the case of $N{=}3$ spin-$1$ scattering centers. Addition of spin $s_1$ and $s_2$ gives $s_{12}{=}0,1,2$. Further coupling of the latter to $s_3$ yields $S{=} 0$ only for $s_{12}{=} 1$. Hence, there is a unique singlet state,  which reads
$|\Psi_{1,3}^-\rangle{=}(|0,1,-1\rangle{+}|1,-1,0\rangle{+}|{-}1,0,1\rangle
-|0,-1,1\rangle{-}|1,0,-1\rangle{-}|{-}1,1,0\rangle)/{\sqrt{6}}$,
where $\ket{m_i}$ is an eigenstate of $\hat{S}_{iz}$
 with quantum number $m_i{=-1,0,1}$.
This state, known also as Aharonov state, was pointed out to be ``{\it so elegant that it must be useful for something}"~\cite{gisin}. Indeed, $|\Psi_{1,3}^-\rangle_{123}$ {can be used as a resource} for the quantum solution to the Byzantine agreement problem~\cite{gisin}. 
{Using our scheme, an Aharonov state can be generated by detecting transmitted $e$ particles that are all scattered (under RCs) by three centers prepared in any of the (product) state components appearing in $|\Psi_{1,3}^-\rangle$}. The associated success probability is $1/6$.

\noindent
{\it Even number of spin-1/2 centers.--} We now consider the general case of $s{=}1/2$ and $N{=}2n$ ($n{\in}\mathbb{N}$). 
Up to a suitable permutation of the labels of the scattering centers, any separable state having $n$ spins in state $|\!\up\rangle$ and $n$ in $|\!\down\rangle$ reads
\begin{equation} \label{product}
|\pi\rangle{=}{\underbrace{|\!\up\up\cdot\cdot\up\rangle}_{n}}\otimes{\underbrace{|\!\down\down\cdot\cdot\down\rangle}_n},
\end{equation}
where the first (second) ket is the state of particles $1,2,..n$ ($n{+}1,n{+}2,..,N$). State $|\pi\rangle$ belongs to the common eigenspace of $\mathbf{S}^2_{12\cdot\cdot\cdot n}$ and $\mathbf{S}^2_{n\!+\!1n\!+\!2 \cdot\cdot\cdot N}$ with quantum numbers $s_{1 2\cdot\cdot\cdot n}{=} s_{n\!+\!1n\!+\!2 \cdot\cdot\cdot N}{=} n/2$. Due to the effective form taken by the scattering potential in~\eq(\ref{Veff-el-N}) under RCs,  these two operators and $\mathbf{S}^2$ are constants of motion. 
As $n/2$ is the maximum value taken by $s_{12\cdot\cdot n}$ and $ s_{n\!+\!1n\!+\!2\cdot\cdot N}$, there is a unique singlet state $|\Psi_{1/2,N}^-\rangle$ belonging to this eigenspace~\cite{notadeg}. This state has the general form \cite{EPAPS}
\begin{equation} \label{ss}
|\Psi_{1/2,N}^-\rangle{=}\frac{1}{\sqrt{n+1}}\sum_{\nu=0}^n(-1)^{n-\nu} |D_n^{(\nu)}\rangle  |D_n^{(n-\nu)}\rangle,
\end{equation}
 where $|D_n^{(0)}\rangle{=}\ket{{\down\cdot\cdot\down}}$, $|D_n^{(n)}\rangle{=}\!\ket{\up{\cdot\cdot}\up}$ and $|D_n^{(0<k<n)}\rangle$ are $n$-particle Dicke states with $k$ excitations ($0\!<\!k\!<\!n$) given by
\begin{equation}
\label{dicke}
\begin{aligned}
|D_n^{(k)}\rangle{=}({1}/{{{B_n^{(k)}}}})^{1/2}\sum_{l} \hat{P}_{l}\, |\!\underbrace{\up{\cdot\cdot}\up}_k\underbrace{\down{\cdot\cdot}\down}_{n-k}\rangle,
\end{aligned}
\end{equation}
where $B_n^{(k)}$ is the binomial coefficient and $\hat{P}_{l}$ is the set of all distinct permutations of $\ket{\up}$ and $\ket{\down}$ states appearing in the representative ket of Eq.~(\ref{dicke}). 
It is immediate to check that, for $n{=}1$, Eq.~(\ref{ss}) reduces to $|\Psi_{1/2,2}^-\rangle{=}(|{\up\down}\rangle\!-\!|{\down\up}\rangle)/\sqrt{2}$. Interestingly, for four scattering centers (${N=4}$), we get~\cite{korea}
\begin{equation}\label{psi-1/2,4}
|\Psi_{1/2,4}^-\rangle{=}\frac{1}{\sqrt{3}}(|\!\up\up\rangle_{12}|\!\down\down\rangle_{34}\!+\!|\!\down\down\rangle_{12}|\!\up\up\rangle_{34}\!-\!|\Psi^+\rangle_{12}|\Psi^+\rangle_{34}),
\end{equation}
where $|\Psi^+\rangle_{jk}{=}(1/\sqrt 2)(|\!\up\down\rangle\!+\!|\!\down\up\rangle)_{jk}~(j,k{=}1,..,4)$ and we have introduced particle labels in order to make the expression self-explanatory. The state in Eq.~(\ref{psi-1/2,4}) has been recently investigated in the context of linear optics settings, where its role in quantum networking protocols (such as secret sharing) has been demonstrated~\cite{qss}. Here, we outline how it can be generated by our scheme. 

{By preparing a state such as $|\!\up\up\rangle_{12}|\!\down\down\rangle_{34}$, we in fact set $s_{12}{=} s_{34}{=}1$. Under RCs, the conservation laws discussed above fully hold and thus  $s_{12}$ and $s_{34}$ are good quantum numbers. Hence, the scattering process of particles $e$ necessarily leaves the state of the system within such subspace. Our protocol thus performs a projective measurement over the only possible singlet state Eq.~(\ref{psi-1/2,4}) with success probability 1/3 (this is $|\langle\up\up\down\down\!|\Psi_{1/2,4}^-\rangle|^2$).}
\begin{figure}
\includegraphics[width=0.45\textwidth]{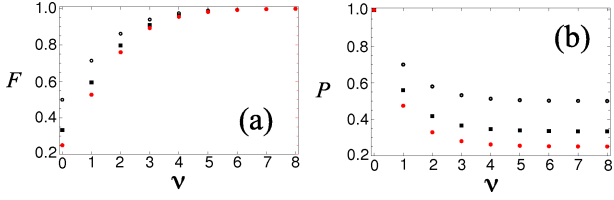}
\caption{(Color online) \textbf{(a)} Fidelity of singlet-state extraction against the number $\nu$ of mobile particles launched into the scattering region. {\bf (b)} The success probability of the scheme. The empty circles show the case corresponding to $N=2$ spin 1/2 scattering centers, the filled squares are for $N=4$, while the filled circles refer to $N=6$. The centers are initially prepared in the state (\ref{product}), whereas spin-spin interaction under RCs occurs via Eq.~(\ref{Vel}) with $J=2$. \label{figura}}
\end{figure}
{This result is naturally generalized to an arbitrary number of centers.
Indeed, by rewriting the state in Eq.~(\ref{product}) as $|\pi\rangle{=}|D_n^{(n)}\rangle|D_n^{(0)}\rangle$ and using Eq.~(\ref{ss}) we conclude that if the centers are prepared in a state having the form $|\pi\rangle$, the singlet state $|\Psi^-_{1/2,N}\rangle$ is extracted with probability $(1\!+\!N/2)^{-1}$.} 

In Fig.~\ref{figura} we show a numerical simulation where a few mobile $e$ particles are scattered from $N{=}2,4$ and $6$ scattering centers. 
To evaluate the performance of the scheme we use the fidelity $F{=}\langle\Psi^-_{1/2,N}|\rho|\Psi^-_{1/2,N}\rangle$ between the target $|\Psi^-_{1/2,N}\rangle$ and the centers' state $\rho$, along with the success probability $P$. Both of these are plotted against the number of detected particles $\nu$. As $\nu$ grows, ${F\rightarrow1}$ while $P\rightarrow(1\!+\!N/2)^{-1}$, in agreement with the analytical results shown above. The coupling strength ${J=2}$ set in Fig.~\ref{figura} yields $F{\gtrsim}0.9$  already after ${\nu=4}$ iterations and regardless of $N$. {Yet, a larger fidelity is obtained} with a lower number of launches if a larger coupling strength is set~\cite{mappaNP}.

\noindent
{\it Optimality for telecloning.--} Here we show that, among the $N$-partite singlet states that can be achieved {using our scheme}, those in Eq.~(\ref{ss}) are ideal resources for optimal ancilla-assisted multi-party telecloning~\cite{murao} from an input scattering center to $n$ ``receiving" ones. This further dresses such class of generable states with practical relevance for multipartite quantum networking and communication. We consider $N{+}1$ spin-1/2 scattering centers with $N{=} 2n$ ($n\in\mathbb{Z}$). One of the centers, denoted by $X$, embodies the {\it input}. Half of the remaining $N$ centers are indexed as $i{=}1,..,n$: particle $1$ is dubbed the {\it port} while centers  $2,..,n$ are used as \emph{ancillae}. The other $n$ particles, labelled as $i{=} n\!+\!1,..,N$, embody the \emph{receivers}. Using the customary jargon in quantum communication, we consider a sending agent, Alice,  who controls the state of $X$ along with that of center 1, namely the port. The task of Alice is to clone the unknown state $|\varphi\rangle_X{=}\alpha\ket{\down}_X\!+\!\beta\ket{\up}_X~(|\alpha|^2{+}|\beta|^2{=}1)$ onto the state of each of the $n$ receivers. In Refs.~\cite{murao,massar} it was shown that an operation that exploits the port to {\it map} state $\ket{\varphi}_X$ onto
\begin{equation} 
\label{optimal}
|\Phi\rangle_{AR}{=}\alpha|\Phi_0\rangle_{AR}+\beta|\Phi_1\rangle_{AR},
\end{equation}
achieves optimal telecloning provided that~\cite{murao} 
\begin{equation}
\label{desiderata}
|\Phi_q\rangle_{AR}\!=\!\sum_{\nu=0}^{n-1}(-1)^{n-\nu}\sqrt{\frac{2 (n-\nu)}{n (n+1)}}\, |A_{\nu_q}\rangle_A\,|D_n^{(\mu_q)}\rangle_R,
\end{equation}
where $q{=}0,1$,  $\mu_{0}{=}\nu_1{+}1{=}n{-}\nu,\,\mu_1{=}\nu_0{=}\nu$. Here, subscript $A$ ($R$) stands for the ensemble of ancillary (receiving) centers and $\{|A_\nu\rangle_A\}$ is a set of $n$ orthonormal states of the ancillae. {In what follows, we show the ideality of the resources achieved by our scheme by showing that}, by means of Eq.~(\ref{ss}), we get a map having the very same structure of Eqs.~(\ref{optimal}) and~(\ref{desiderata}). 

We consider the initial overall state $|\Psi_0\rangle_{X\tilde{A}R}{=} |\varphi\rangle_X\otimes|\Psi_{1/2,N}^-\rangle_{\tilde{A}R}$ with $\tilde{A}{=}1{\cup}{A}$. The first step of the protocol requires that Alice implements a Bell measurement of the joint state of spins $X$ and the port $1$. As proven in Ref.~\cite{mappaNP} and explicitly used in~\cite{teleportic}, the same interaction model and iterative scheme discussed here for singlet-state extraction can be used to implement an efficient Bell measurement.  Without affecting the generality of our discussions, we will focus on projections onto the Bell state 
$|\phi^+\rangle_{X1}{=}(\ket{\up\up}_{X1}\!+\!\ket{\down\down}_{X1})/\sqrt 2$. We then find convenient to decompose state $|\Psi_{1/2,N}^-\rangle_{\tilde{A}R}$ as 
$|\Psi_{1/2,N}^-\rangle_{\tilde{A}R}{=}|\Psi_0\rangle_{\tilde{A}R}+|\Psi_1\rangle_{\tilde{A}R}$
with
\begin{equation}
\label{psi0psi1}
\begin{aligned}
|\Psi_0\rangle_{\tilde AR}&{=}\frac{(-1)^n|D_n^{(0)},D_n^{(n)}\rangle_{\tilde AR}+|D_n^{(n)},D_n^{(0)}\rangle_{\tilde AR}}{\sqrt{n+1}},\\
|\Psi_1\rangle_{\tilde AR}&{=}\frac{1}{\sqrt{n+1}}\sum_{\nu=1}^{n-1}(-1)^{n-\nu}\,|D_n^{(\nu)},D_n^{(n-\nu)}\rangle_{\tilde AR}.
\end{aligned}
\end{equation}
It is easy to see that
${}_{X1}\langle\phi^+|\varphi,\Psi_0\rangle_{X\tilde AR}$ changes the state of the ensemble $A{\cup}R$ into $[\alpha(-1)^n|D_{n-1}^{(0)},D_{n}^{(n)}\rangle_{AR\!}+\beta|D_{n-1}^{(n-1)},D_{n}^{(0)}\rangle_{AR}]/\sqrt{2(n+1)}$, where we have used that $\langle\phi^+|{\down,D_n^{(0,n-1)}}\rangle{=}|D_{n-1}^{(0)}\rangle/\sqrt{2}$ and $\langle\phi^+|{\up,D_n^{(n)}}\rangle{=}|D_{n-1}^{(n-1)}\rangle/\sqrt{2}$.
As for $_{X1}\langle\phi^+|\varphi,\Psi_1\rangle_{XAR}$, Eq.~(\ref{psi0psi1}) shows that we need $\langle\phi^+|{\down,D_n^{(\nu)}}\rangle=\sqrt{{(n-\nu)}/{2n}}|D_{n-1}^{(\nu)}\rangle$ and $\langle\phi^+|{\up,D_n^{(\nu)}}\rangle=\sqrt{{\nu}/{2n}}\,|D_{n-1}^{(\nu-1)}\rangle$. 

Putting all this together, we find that the initial state $|\Psi\rangle_{X\tilde AR}$ is transformed into $|\phi^+\rangle_{X1}\otimes|\tilde{\Phi}\rangle_{AR}$, where
\begin{equation}
\label{final2}
\begin{aligned}
|\tilde{\Phi}\rangle_{AR}&{=}\alpha\left[\frac{\,(-1)^n|D_{n-1}^{(0)},D_{n}^{(n)}\rangle_{AR}}{\sqrt{(n+1)/2}}+\sum_{\nu=1}^{n-1}\frac{\chi_{n-\nu}|D_{n-1}^{(n)}, D_n^{(n-\nu)}\rangle_{AR}}{\sqrt{n\,(n+1)/2}}\right]\\
&+\beta\left[\frac{|D_{n-1}^{(n-1)},D_n^{(0)}\rangle_{AR}}{\sqrt{(n+1)/2}}+\sum_{\nu=1}^{n-1}{\frac{\chi_{\nu}|D_{n-1}^{(n-\nu-1)},D_{n}^{(\nu)}\rangle_{AR}}{\sqrt{n\,(n+1)/2}}}\right]
\end{aligned}
\end{equation}
with $\chi_{\eta}\!=\!(-1)^{\eta}\sqrt{n-\nu}$. Eq.~(\ref{final2}) has the same form as \eq (\ref{optimal}), thus demonstrating the ideal nature of the resource produced by the scheme for multipartite singlet-state extraction proposed here. By changing our choice of the Bell state onto which Alice performs her projections, only local operations at the receiving sites will be needed in order to retrieve Eq.~(\ref{final2}). 

\noindent
{\it Generation of GHZ and W states.--} In order to illustrate the flexibility of our scheme for the generation of a variety of interesting and useful multipartite entangled resources, we now address a significant case. We will see that a judicious {choice of the initial  product state} and suitable post-selection operations are key to the achievement of relevant multipartite states. 

We consider the case of ${N\!=\!4}$ spin-$1/2$ centers whose spins are coupled together
according to the \emph{(12){-}(34)} scheme. The \emph{triplet} subspace of quantum numbers $s_{12}{=} s_{34}{=} s_{1234}{=}1$ is spanned by the states
$|M{=}1\rangle_{1234}{=}|D^{(3)}_{4}\rangle$, $|M{=}0\rangle_{1234}{=}(|{\up\up\down\down}\rangle{-}|{\down\down\up\up}\rangle)/\sqrt{2}$, $|M{=-1}\rangle_{1234}{=}|D^{(1)}_4\rangle$~\cite{korea}.
Here, $M$ is the azimuthal quantum number associated with $\hat{S}_z{=}\sum^4_{i=1}\hat{S}_{i,z}$. Evidently, state $\ket{M=0}$ is of the GHZ class~\cite{GHZ}, while $|{M{=}{1}}\rangle$ ($|{M{=}{-1}}\rangle$) is a four-particle Dicke states with three (one) excitations, {\it i.e.} a W state~\cite{Dur}. These non-equivalent classes of entangled states have been the objects of extensive endeavors for  their experimental production and use in quantum communication protocols.

If we now consider a fifth scattering center with ${s_5{=}1}$, a unique singlet state arises in the eigenspace with $s_{12}{=} s_{34}{=}s_{1234}{=}1$, reading
\begin{equation} \label{psi}
|\Psi\rangle{=}({|{-1},1\rangle_{12345}{-}|0,0\rangle_{12345}{+}|1,{-1}\rangle_{12345}})/{\sqrt{3}},
\end{equation}
where we have used the notation $\ket{M,m_5}_{12345}{=}\ket{M}_{1234}\otimes\ket{m_5}_{5}$.
Depending on the outcome of a measurement performed over the computational basis $\{|1\rangle_5,|0\rangle_5, |\!-\!1\rangle_5\}$ of center 5, the four spin-$1/2$ particles are projected either onto a GHZ state with probability $1/3$, or onto one of the two W states, each obtained with probability $1/3$. Remarkably, \emph{any} outcome of such measurement yields an exploitable entanglement resource. State~(\ref{psi}) can be generated with probability $1/6$ using the proposed extraction scheme upon preparation of the centers' product state $|\!\up\up\down\down\rangle_{1234}|0\rangle_5$~\cite{M}.

\noindent
{\it Conclusions.--}  We have proposed a physical model for the systematic generation of $N$-partite states for quantum networking. The physics behind our idea has been clearly interpreted using the formalism of coupled angular momenta and the symmetries arising from an effective spin-spin scattering mechanism~\cite{prl,mappaNP,yuasa}. We have shown that, among other resources, our scheme is able to produce scalable multipartite singlet states for the performance of optimal telecloning. To the best of our knowledge, this is the first demonstration of the existence of a class of generable mutipartite states having such features. By highlighting a few interesting applications that the variety of states we address can have, we have dressed our proposal of pragmatic relevance. Physically, our scheme is based on only rather weak requirements and exploits well-spaced, stationary elements of the register, which are major practical advantages. 

Very recently, it has been shown \cite{daniel} that in the case of two spin-1/2 centers the singlet-extraction 
protocol that we have used here can be modified so as to make it deterministic. Extension of such procedure to the various situations addressed in this paper is under current investigations.

\noindent
{\it Acknowledgments.--}  We thank Y. Omar and M. Weber for discussions. SB acknowledges support from the EPSRC, the Royal Society and the Wolfson Foundation. DEB thanks the NSF of Singapore and QNET. GMP acknowledges EUROTECH. MP is supported by EPSRC (EP/G004579/1).

\appendix

\section{Supplementary material}

Here, we provide a rigorous and technical proof of Eq.~(4)  of the main paper, which decomposes the singlet state of an even number $N$ of spin-$1/2$ particles in terms of Dicke states with $k$ excitations.  The crucial property of the singlet state $\!|\Psi_N^-\rangle$ is that it is a simultaneous eigenstate of $\mathbf{S}_{1\cdot\cdot\cdot n}^2$, $\mathbf{S}_{n+1\cdot\cdot N}^2$ and $\mathbf{S}^2$ with quantum numbers $s_{1\cdot\cdot n}\ug n/2$, $s_{n+1\cdot\cdot N}\ug n/2$ and $S\ug0$, respectively. In terms of the angular-momentum states of the two subsystems of $n$ spins into which the system of $N\!=\!2n$ scattering centers is partitioned, we have
\begin{equation} \label{ss2}
\!|\Psi_{1/2,N}^-\rangle\ug\!\sum_{\mu=-n/2}^{n/2} c(n/2,\mu,n/2,-\mu; 0,0) \,|n/2,\mu\rangle|n/2,-\mu\rangle,
\end{equation}
where $|n/2,\mu\rangle$ are spin-$n/2$ states with azimuthal quantum number $\mu{=}\nu{-}n/2$, where $\nu$ is the number of particles in $|{\up}\rangle$. Moreover, we have introduced the numbers
\begin{equation} \label{CG}
c(j_1,m_1,j_2,m_2; j, m)\ug \langle j_1,m_1|\langle j_2, -m_2|j,m\rangle,
\end{equation}
which are standard Clebsch-Gordan coefficients for the coupling of angular momenta~\cite{aam}. 
Using this and the relation,
\begin{equation}
c(j_1,m_1,j_2,m_2; 0, 0)\ug\frac{(-1)^{j_1\!-\!m_1}}{\sqrt{(2j_2\!+\!1)}}\delta_{j1,j2}\delta_{m_1,-m_2}
\end{equation}
\eq (\ref{ss2}) takes the form
\begin{equation} 
\label{ss3}
|\Psi_{1/2,N}^-\rangle\ug\frac{1}{\sqrt{n+1}} \sum_{\nu\ug 0}^n (-1)^{n-\nu}  \,|n/2,\mu\rangle|n/2,-\mu\rangle.
\end{equation}
Therefore, in order to demonstrate Eq.~(4), our task reduces to show that $|n/2,\mu\rangle\ug \,|D_{n}^{(\nu)}\rangle$ for any $n$ and $\mu\ug-n/2,..,n/2$. We prove such identity by induction. 

For $n\ug1$, we trivially have that $|{\up}\rangle\ug|1/2,1/2\rangle\ug|\!D_1^{(1)}\rangle$.  We will thus  assess the case of a subsystem of $n+1$ particles, by assuming that the relation holds for $n$. For $n+1$ spins and $\nu\ug0,n\!+\!1$ the relation is true given that $|D_{n+1}^{(0)}\rangle\ug|D_{n}^{(0)}\rangle|\!\down\rangle$ and  $|D_{n+1}^{(n+1)}\rangle\ug|D_{n}^{(n)}\rangle|\!\up\rangle$. To prove it also for $0\!<\!\nu\!<\!n\!+\!1$, we first notice that the definition of the Dicke states implies
\begin{equation} \label{rule}
|D_{n+1}^{(\nu)}\rangle\ug\!\frac{\sqrt{B_{n}^{(\nu-1)}}|D_n^{(k-1)}\rangle|\!\up\rangle\!+\!\sqrt{B_{n}^{(\nu)}}|D_n^{(\nu)}\rangle|\!\down\rangle}{\sqrt{B_{n+1}^{(\nu)}}},
\end{equation}
with $B^{(\nu)}_n$ the Binomial coefficient. On the other hand, by coupling an effective spin-$n/2$ particle (with quantum number $s_{12\cdot\cdot\cdot n}\ug n/2$) to a further spin-$1/2$ one, so as to achieve a system of $n+1$ particles with  $s_{12\cdot\cdot\cdot n\!+\!1}\ug(n\!+\!1)/2$, we get
\begin{equation} 
\label{addition}
\begin{aligned}
&|\frac{n+1}{2}, \mu'\rangle\ug c\left(\frac{n}{2},\mu'\!-\!\frac{1}{2},\frac{1}{2},\frac{1}{2};\frac{n\!+\!1}{2},\mu'\right)|\frac{n}{2},\mu'\!-\!\frac{1}{2}\rangle|\!\up\rangle\\
&+ c\left(\frac{n}{2},\mu'\!+\!\frac{1}{2},\frac{1}{2},-\frac{1}{2};\frac{n\!+\!1}{2}, \mu'\right)|\frac{n}{2},\mu'\!+\!\frac{1}{2}\rangle|\!\down\rangle.
\end{aligned}
\end{equation}
The two Clebsch-Gordan coefficients appearing in \eq (\ref{addition}) can be arranged in the compact form
\begin{eqnarray}
 \label{CG2}
c\left(\frac{n}{2},\mu'\!\pm\!\frac{1}{2},\frac{1}{2},\mp\frac{1}{2};\frac{(n\!+\!1)}{2}, \mu'\right)\propto\sqrt{\frac{{n+1\mp 2 \mu'}}{{{2 \,(n+1)}}}},
\end{eqnarray}
where the proportionality sign holds because of an unimportant phase factor common to both the Clebsch-Gordan coefficients which would simply represent a global phase for $\ket{(n+1)/2,\mu'}$. 
\eq (\ref{CG2}) thus entails
\begin{eqnarray}
c\left(\frac{n}{2},\mu'\!\pm\!\frac{1}{2},\frac{1}{2},\mp\frac{1}{2};\frac{(n\!+\!1)}{2}, \mu'\right)=\sqrt{\frac{B_n^{(\mu'\pm1/2+n/2)}}{B_n^{(\mu'+n/2+1/2)}}}.
\end{eqnarray}
In~\eq(\ref{addition}) we can replace $|n/2,\mu'\!-\!1/2\rangle$ ($|n/2,\mu'\!+\!1/2\rangle$) with $|D_n^{(\mu'\!-\!1/2\!+\!n/2)}\rangle$ ($|D_n^{(\mu'\!+\!1/2\!+\!n/2)}\rangle$) due to the hypothesis that the identity $|n/2,\mu\rangle\ug \,|D_{n}^{(\nu)}\rangle$ is valid for a system of $n$ spins. Using such replacement along with \eq (\ref{CG2}) and noticing that the number of $\up$-polarized particles in $|(n+1)/2,\mu'\rangle$ is $\nu'\ug\mu'\!+\!(n\!+\!1)/2$, our thesis is proved upon comparison of \eq (\ref{rule}) for $k{=}\nu'$ and \eq (\ref{addition}).

\end{document}